\documentclass[12pt,english,fleqn,twoside]{article}
\usepackage[T1]{fontenc}
\usepackage[latin1]{inputenc}
\usepackage{babel}
\usepackage{graphics}

\makeatletter

\providecommand{\LyX}{L\kern-.1667em\lower.25em\hbox{Y}\kern-.125emX\@}

\usepackage{verbatim}

\usepackage{espcrc1}
\usepackage{graphicx}

\makeatother
\begin{document}

\title{Continuum effects in reactions involving weakly bound nuclei}

\author{A. M. Moro\address[IST]{Departamento de F\' {\i}sica, 
Instituto Superior T\'{e}cnico, 1049-001 Lisboa, Portugal}, 
R. Crespo\addressmark[IST], F. Nunes\address{Universidade Fernando Pessoa, 
Pra\c{c}a 9 de Abril, 4200 Porto, Portugal} and 
I. J. Thompson\address{Departament of Physics, University of Surrey, Guildford GU2 5XH, UK}}

\maketitle

\begin{abstract}
In this contribution we investigate
the role of  the continuum spectrum
in reactions of astrophysical interest.
In particular, the influence of these coupling effects in the description
of elastic and transfer reactions is discussed. 
We examine the validity
of the Distorted Wave Born Approximation (DWBA) as a tool to extract
the astrophysical \emph{\( S_{17}(0) \)} factor from the 
transfer reaction 
$^{14}$N($^{7}$Be,$^{8}$B)$^{13}$C. 
For this purpose,
we present calculations for this  reaction 
comparing the DWBA method with the
CDCC-Born approximation, 
in which continuum effects are explicitly incorporated. 
\end{abstract}

\section{INTRODUCTION}
In recent years, much of the effort  in nuclear physics has been devoted 
to the study of nuclear reactions of astrophysical interest.
The direct
measurement of these reactions at the relevant energies (tens of keV or less) 
is still beyond the capability of the existing facilities. 
From the theoretical point of view, the description of these
reactions can be substantially simplified by noting that, at these low
energies, the reaction mechanism is only sensitive to the tail of the
wave-function for the participants.  Thus, the rate of the capture reaction
$a(p,\gamma )b$
is essentially determined by the asymptotics of the 
single-particle overlap $(b|a)$ which, in turn, is proportional
to a Whittaker function. The proportionality constant, known as the asymptotic
normalization coefficient (ANC), 
determines unambiguously the capture cross 
section at zero energy.

This feature constitutes the basis  
of the ANC method, which  extracts the ANC for the
overlap  $(b|a)$  from the analysis of a transfer reaction 
$A(a,b)B$. As an important prerequisite, the selected 
transfer reaction has to be peripheral. Under this condition, the DWBA amplitude for the transfer process
is proportional to  the product
of the ANC's $C_{ba}$ and  $C_{BA}$. Thus, by comparison 
of the calculated and 
measured transfer cross section, 
it is possible to extract the value for  $C_{ba}$, provided that
the ANC  $C_{BA}$ is already known.
Due to the approximations involved in the DWBA, the 
applicability of the method requires that
the transfer  occurs in one-step and that the entrance and
exit channels are well described by phenomenological optical potentials~\cite{Fer00}.

When weakly bound nuclei are involved, the transfer mechanism might
occur through continuum states. Thus, it is not obvious that these multi-step
processes are properly accounted for by an optical model
potential, as it is assumed in the DWBA approach. In 
these situations, it may result
more adequate to treat explicitly the couplings involving the 
continuum states. A suitable formalism to include these effects
is the CDCC-BA approach. 
Here, the partition that contains the weakly bound nucleus is
described within a model space that includes the continuum spectrum of 
that nucleus, by means of a procedure of discretization. 
Formally, this is  done by 
replacing the total wavefunction in the exact transition
amplitude by a solution of a Continuum 
Discretized Coupled Channels calculation~\cite{Mor02b}.

Recently, the reaction \( ^{14} \)N(\( ^{7} \)Be,\( ^{8} \)B)\( ^{13} \)C
at a bombarding energy of 85 MeV has been measured at the Texas A\&M
Cyclotron facility in order provide an indirect determination of the 
astrophysical $S_{17}$ factor, using the ANC for the overlap 
(\( ^{7} \)Be,\( ^{8} \)B). This ANC was extracted by analyzing 
the experimental data within the  DWBA approach \cite{Azh99}. Since 
the \( ^{8} \)B nucleus is weakly bound 
(\( \epsilon _{0}=0.137 \)
MeV), transfer could proceed
through the \( ^{8} \)B continuum states and hence this approach 
could be inappropriate. In this work, we aim to 
evaluate the effect of the \( ^{8} \)B continuum in the cited transfer
reaction. For this purpose, we have performed an  analysis
of the data using the CDCC-BA approach and comparing our results with those
obtained in the DWBA method.
Since the DWBA
and CDCC-BA approaches differ essentially in the description
of the exit channel, we start by analyzing 
the elastic scattering \( ^{8} \)B+\( ^{13} \)C.


\begin{figure}
{\centering \resizebox*{0.5\columnwidth}{!}{\includegraphics{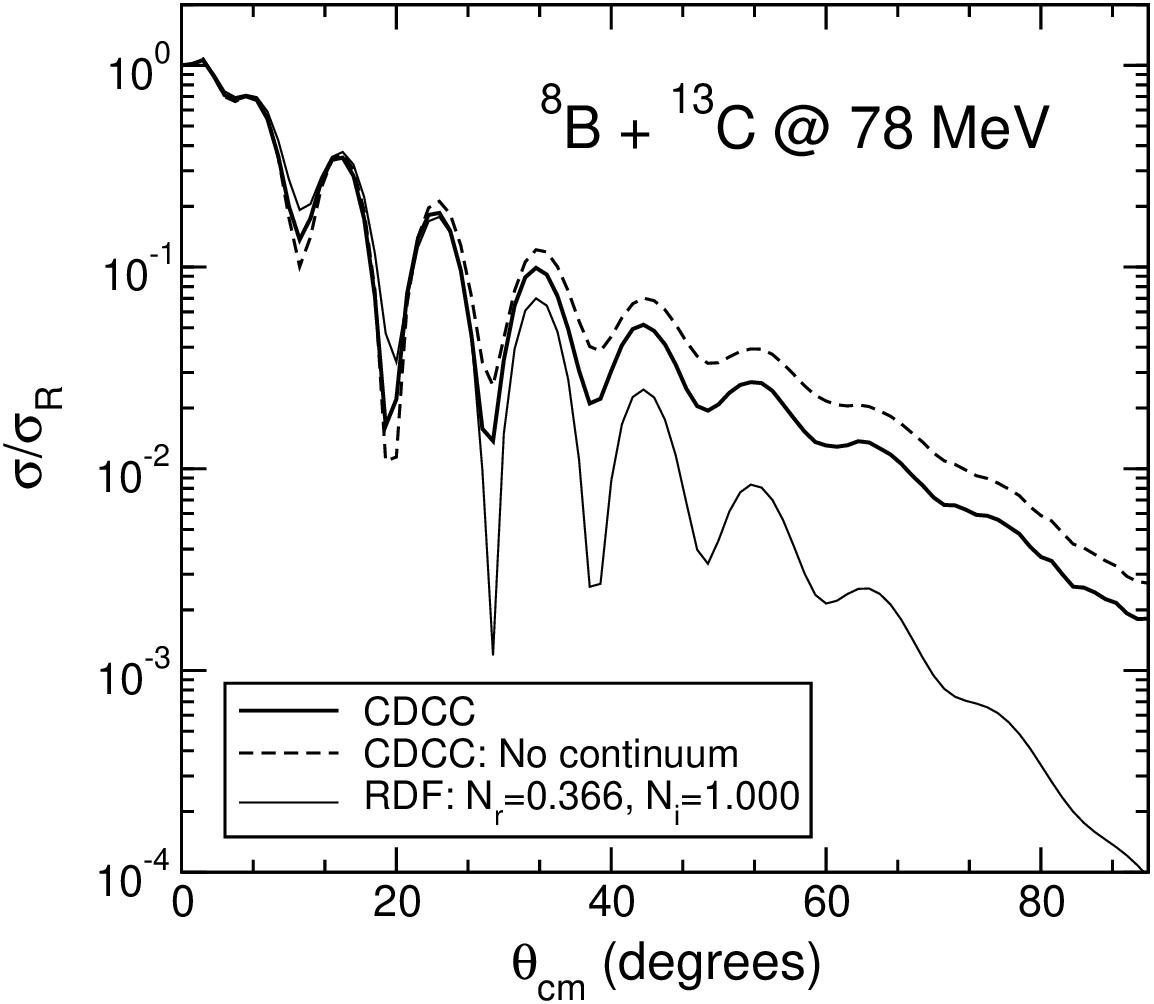}} \par}
\caption{\label{Fig:b8c13_elas}Angular distribution for the differential 
elastic cross sections
calculated with the CDCC method
(thick solid line) and RDF potential (thin
solid line). The dashed line is the CDCC calculation without
continuum couplings.}
\end{figure}

\section{ANALYSIS OF THE ENTRANCE AND EXIT CHANNELS}

In the DWBA analysis of Azhari et al~\cite{Azh99}, the
elastic
scattering  $^{7}$Be+$^{14}$N was described using a double 
folding potential, with an average normalization constant 
extracted from the analysis of reaction with nearby nuclei \cite{Trac00}.
As shown in~\cite{Azh99}, this renormalized
double folding  (RDF) potential
reproduces very well the experimental data for the
elastic scattering \( ^{7} \)Be+\( ^{14} \)N. In the present analysis, 
we have also adopted this potential to describe the entrance channel in 
both the DWBA and CDCC-BA calculations.

The lack of experimental
data for the reaction \( ^{8} \)B+\( ^{13} \)C, due to the small
intensities of the present beams of \( ^{8} \)B, makes the optical
potential for the exit channel more uncertain. In ~\cite{Azh99}, 
the RDF potential was used. Note
that this ambiguity is absent in the CDCC approach, 
since the \( ^{8} \)B+\( ^{13} \)C
interaction is described in terms of the cluster-target interactions.
In the present work, we have assumed that the \( ^{8} \)B nucleus consists
of a inert \( ^{7} \)Be core plus a valence proton. The \( ^{7} \)Be-\( p \)
continuum is discretized into \( N=10 \) bins, uniformly spaced in
energy, and up to \( \epsilon _{\max}= \) 9~MeV. We took into account
\( s, \) \( p \), and \( d \) partial waves for the relative motion.
Details of binding and optical potentials used in the calculations
can be found in~\cite{Mor02b} and references therein.


In Fig.~\ref{Fig:b8c13_elas} we show the calculated elastic scattering
for \( ^{8} \)B+\( ^{13} \)C at \( E_{lab} \)=78 MeV. The thin
solid line is the calculation with the RDF potential, whereas the thick solid
line is the full CDCC result. We find a good agreement between both
calculations at small angles, but they differ significantly at larger
angles, in the region where continuum effects become relevant. Although
experimental data would be required to clarify the accuracy of both
methods, this result cast doubt on the validity of global parameterizations
when weakly bound nuclei are involved. Also shown in this figure (dashed line)
is the CDCC calculation without continuum couplings, 
which corresponds just to a cluster folding
calculation. By comparing this calculation with the full CDCC,
we find that continuum effects become important for scattering
angles above  30 degrees. Although not shown in this figure, we found also 
that the calculation with only g.s.-continuum 
couplings gives a result very close
to the full CDCC, indicating that continuum-continuum effects are
small in this reaction. 

\section{THE TRANSFER REACTION}
Although the DWBA and CDCC-BA amplitudes
differ in the description of the exit channel, they involve the same
transition operator. In prior form, this operator is  $V_{p,14N}+V_{rem}$, 
where $V_{rem}=U_{7Be,13C}-U_{7Be,14N}$ 
is commonly referred as remnant term.

In order to allow a reliable comparison between
the DWBA and CDCC-BA calculations, the
same core-core ( \( ^{7} \)Be + \( ^{13} \)C) 
interaction was used in  both methods.
In particular, we took the  \( ^{7} \)Li+$^{13}$C optical potential
derived in~\cite{Trac00}.
The \( (^{14} \)N|\( ^{13} \)C) overlap
was described using a  \( p_{1/2} \) single-particle wave-function, 
with spectroscopic
factor 0.604~\cite{Trac98}. For the (\( ^{8} \)B|\( ^{7} \)Be) overlap,
we assumed a pure \( p_{3/2} \) configuration with spectroscopic
factor \( S_{p_{3/2}}=0.733 \). The CDCC calculations had to be multiplied
also by the geometric factor 5/16, that arises from the fact that
our model assumes that \( ^{7} \)Be has zero spin~\cite{Mor02b}. 

\begin{figure}[thb]

\begin{minipage}[t]{75mm}
{\centering \resizebox*{!}{0.8\columnwidth}{\includegraphics{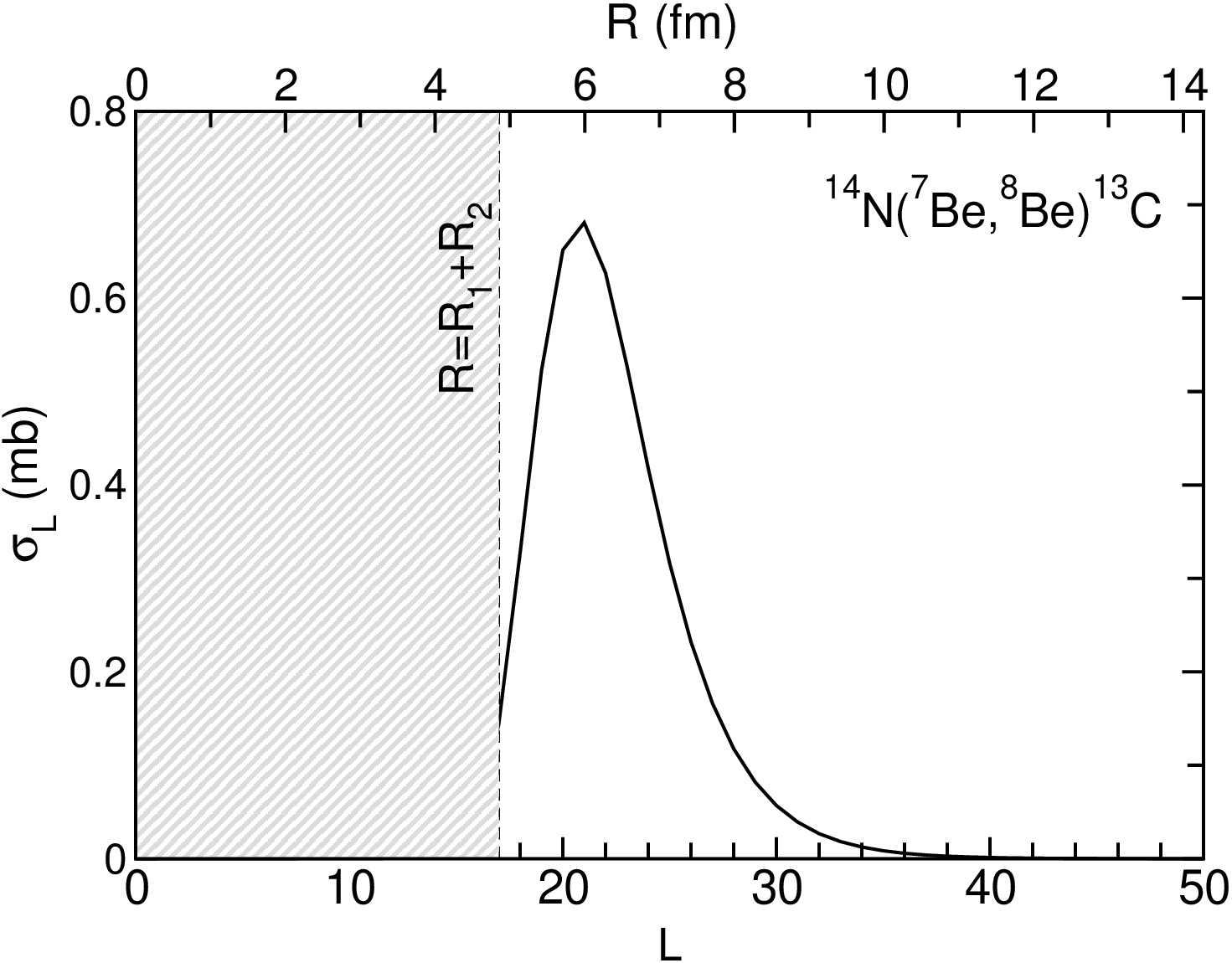}} \par}
\caption{\label{Fig:peripheral}Transfer cross section versus the total angular
momentum. The shaded region corresponds to \protect\( R<R_{1}+R_{2}\protect \)
where \protect\( R_{1}\protect \) and \protect\( R_{2}\protect \)
are estimates for the radii of the colliding nuclei. }

\end{minipage}
\hspace{\fill}
\begin{minipage}[t]{75mm}
{\centering \resizebox*{!}{0.73\columnwidth}{\includegraphics{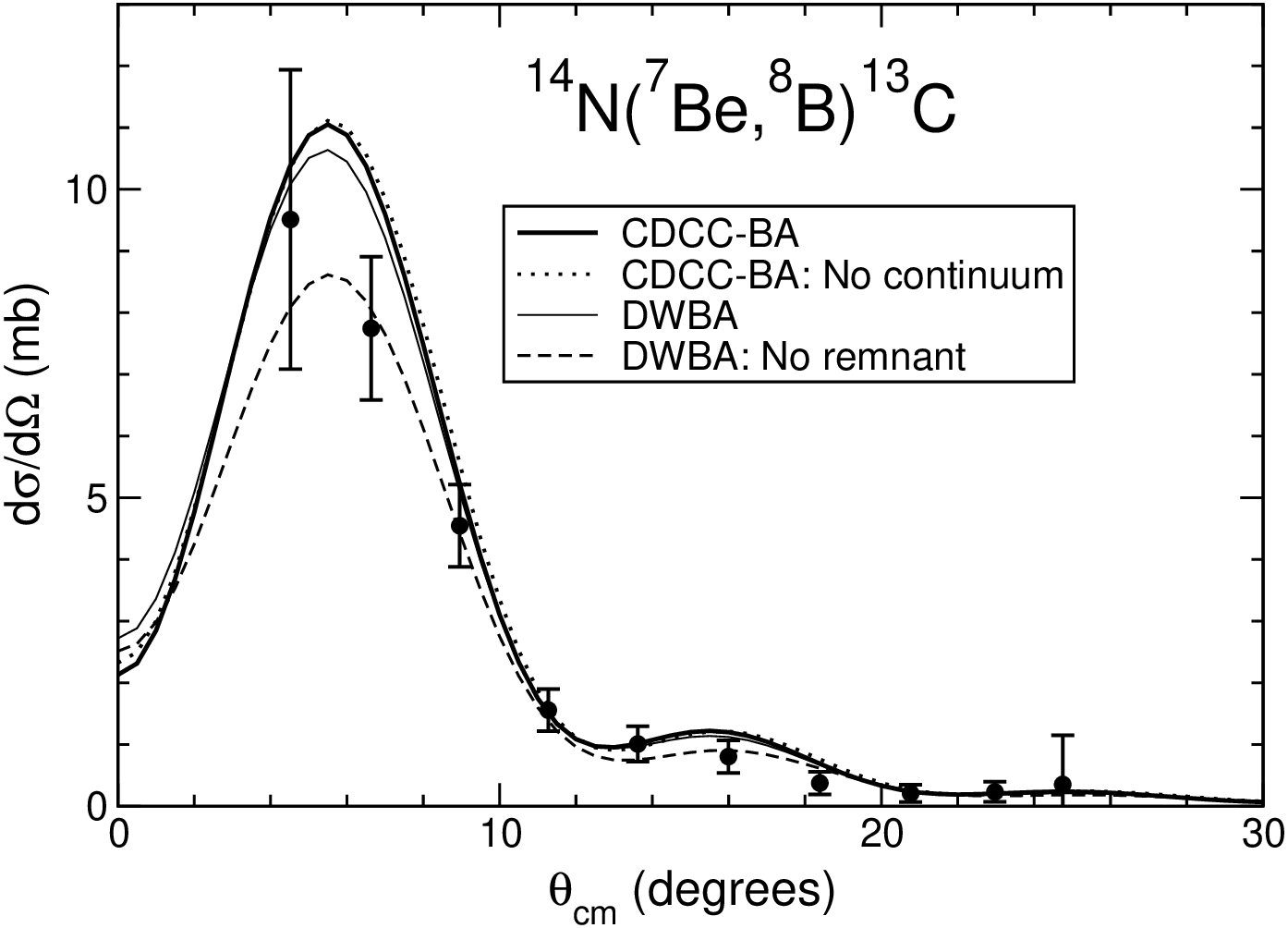}} \par}
\caption{\label{Fig:transfer}
Experimental and calculated transfer cross sections
for the reaction  $^{14}$N($^7$Be,$^8$B)$^{13}$C   at 84 MeV.}

\end{minipage}

\end{figure}

In Fig.~\ref{Fig:peripheral} we present the DWBA transfer cross
sections as a function of the total angular momentum, \( L \), and the impact
parameter \( R=L/k_0 \), where \( k_0 \) is the incident linear momentum. It
is observed that the transfer occurs mainly at distances \( R>R_{1}+R_{2} \),
where \( R_{1} \) and \( R_{2} \) are an estimate for the radii
of the projectile and target. This confirms the peripheral character
of the present reaction, which is a prerequisite for the
applicability of the ANC method.

In Fig.~\ref{Fig:transfer} the experimental and calculated cross
sections for the transfer reaction are shown. The 
thin solid line and the thick solid line correspond to the 
DWBA and CDCC-BA calculations, respectively. 
The dotted line is
a CDCC-BA calculation without continuum couplings. We see that 
the three curves give very
similar results, indicating that in this reaction continuum effects
are not important and thus supporting the validity of the DWBA as
a reliable method to extract the ANC for 
the overlap \( (^{8} \)B|\( ^{7} \)Be). Also shown in this figure,
is the DWBA calculation without remnant term in the 
transition amplitude. The sizable difference with the exact DWBA calculation
shows that the full DWBA transition operator is essential for a quantitative analysis of this
reaction.

\begin{figure}[t]
{\centering \resizebox*{0.9\columnwidth}{!}{\includegraphics{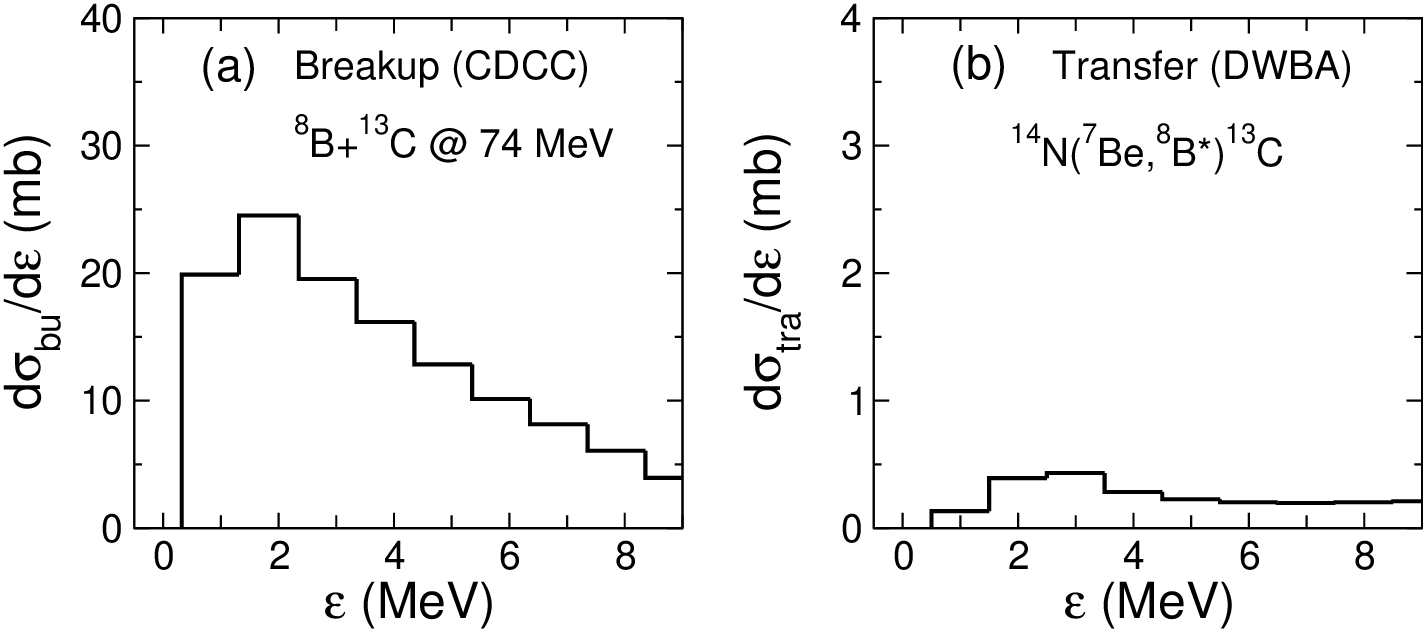}} \par}
\caption{\label{Fig:breakup-trans}
(a)  Breakup cross 
section, calculated in the CDCC approach, for the scattering $^8$B+$^{13}$C at 74 MeV.
(b)  DWBA cross section leading to $^8$B continuum states for 
the reaction  $^{14}$N($^7$Be,$^8$B)$^{13}$C at 84 MeV. }
\end{figure}

In order to understand the reason why continuum effects have such a small effect on 
the transfer cross sections we compare in Fig.~\ref{Fig:breakup-trans}
the DWBA cross section  for
the process $^{14}$N($^7$Be,$^8$B)$^{13}$C with the breakup cross section (calculated with CDCC) 
for the exit channel  $^8$B+$^{13}$C.
In order to have enhanced continuum effects in  $^{14}$N($^7$Be,$^8$B)$^{13}$C, both 
the continuum excitation couplings and the transfer couplings need to be strong.
In Fig.~\ref{Fig:breakup-trans}(a) we see that, as expected, the 
first condition is satisfied in this reaction, since there is an
important breakup probability to continuum states. However, 
transfer couplings to continuum states 
are comparatively much smaller (partially due to the unfavourable Q value), as 
becomes evident in Fig.~\ref{Fig:breakup-trans}(b). Thus, the multi-step process consisting on 
transfer to the  $^7$Be+$p$ continuum and
deexcitation to the  $^8$B ground state are supressed.

\section{CONCLUSIONS}

In summary, we have studied the effect of \( ^{8} \)B two--body continuum
in the transfer reaction \( ^{14} \)N(\( ^{7} \)Be,\( ^{8} \)B)\( ^{13} \)C.
For this purpose, we have compared the standard DWBA method, with the 
CDCC-BA approach. The later takes into account ground state--continuum couplings and
continuum--continuum couplings for the
$^{8}$B continuum.

 We find that the elastic scattering predicted by the CDCC calculation for
the exit channel (\( ^{8} \)B+\( ^{13} \)C)
agrees very well with the renormalized double folding potential used
in~\cite{Trac00} at small angles, but differs significantly at
larger angles. However, these discrepancies do not seem to have any
effect in the forward angle transfer cross sections. Therefore, our calculations
confirm the applicability of the DWBA for the present reaction and, hence,
the value of the ANC extracted in~\cite{Azh99}.

\paragraph{ACKNOWLEDGMENTS}
This work has been partially supported by Fundaç\~ao para a Ciência
e a Tecnologia (F.C.T.) under Grant No. SAPIENS/36282/99 and EPSRC
under Grant No. GR/M/82141. One of the authors, A.M.M., acknowledges
a F.C.T. post-doctoral grant. 


\end{document}